\documentclass[a4paper,12pt]{article}
\usepackage{epsfig,psfrag}
\usepackage{citesort}
\date{\today}
\newcommand{\bmat}{\begin{array}}
\newcommand{\emat}{\end{array}}
\newcommand{\be}{\begin{equation}}
\newcommand{\ee}{\end{equation}}
\newcommand{\bea}{\begin{eqnarray}}
\newcommand{\eea}{\end{eqnarray}}

\def\lsim{\raise0.3ex\hbox{$\;<$\kern-0.75em\raise-1.1ex\hbox{$\sim\;$}}}
\def\gsim{\raise0.3ex\hbox{$\;>$\kern-0.75em\raise-1.1ex\hbox{$\sim\;$}}}

 \normalsize

\begin{document}

\date{\today}

\renewcommand{\thefootnote}{\fnsymbol{footnote}} %\begin{titlepage}
%\rightline{hep-ph/0510xxx}
\rightline{\today} \vspace{.5cm}

\begin{center}
{\large \textbf{Supersymmetry and CP violation in $|\Delta S|=1$ $\tau$
decays} }
\end{center}

\vspace{.1cm}

\begin{center}
D. Delepine$^1$, G. Faisl$^{2}$, S. Khalil$^{2,3}$, and G. L\'opez Castro$^4$%
\\[0pt]

\vspace{.3cm} \emph{$^1$ Instituto de F\'{\i}sica, Universidad de
Guanajuato, 37150 Le\'on, Guanajuato, M\'exico}\\[0pt]
\emph{$^2$ Ain Shams University, Faculty of Science, Cairo 11566, Egypt.}\\[%
0pt]
\emph{$^3$ Center for Theoretical Physics, British University in Egypt,
Shorouk city, Cairo, 11837, Egypt.}\\[0pt]
\emph{$^4$ Departamento de F\'\i sica, Cinvestav, Apdo. Postal 14-740, 07000
M\'exico D.F, M\'exico}\\[0pt]
\end{center}

\vspace{.5cm} \abstract{We compute the SUSY effective hamiltonian
that describes the $|\Delta S|=1$ semileptonic decays of tau
leptons. We provide analytical expressions for supersymmetric
contribution to $\tau \to u \bar{s} \nu_{\tau}$ transition in mass
insertion approximation. We show that SUSY contributions may
enhance the CP asymmetry of $\tau \to K \pi \nu_{\tau}$  decays by
several orders of magnitude than the standard model expectations.
However, the resulting asymmetry is still well below the current
experimental limits obtained by CLEO collaborations. We emphasize
that measuring CP rate asymmetry in this decay larger than
$10^{-6}$ would be a clear evidence of physics beyond the
supersymmetric extensions of the standard model.}

%%%%%%%%%%%%%%%%%%%%%%%%%%%%%%%%%%

\section{{\protect\large \textbf{Introduction}}}

Strangeness-changing tau lepton decays play an important role in
testing the dynamics of $|\Delta S|=1$ weak interactions
\cite{Davier:2005xq} . Determination of basic parameters of the
Standard Model (SM) and tests of fundamental symmetries can be
done using such tau decays. For instance, measurements of the
spectral functions of tau decays into strange mesons have been
used recently to obtain information on the mass of the strange
quark and on the $V_{us}$ CKM matrix element \cite{Gamiz:2004ar}.
Furthermore, searches for CP violation effects in the double
kinematical distributions of $\tau^{\pm} \to K_S
\pi^{\pm}\nu_{\tau}$ decays have been performed recently by the
CLEO Collaborations \cite{Bonvicini:2001xz}. These exclusive
decays can be used to provide further tests on the violation of
the CP symmetry \cite%
{Bonvicini:2001xz,Kuhn:1996dv,Bigi:2005ts,Delepine:2005tw}. A `known' CP
rate asymmetry of $O(10^{-3})$ has been pointed out to exist between $\tau^-
\to K_{L,S} \pi^-\nu_{\tau}$ and their CP conjugate decays \cite{Bigi:2005ts}%
. On the other hand, within the SM, the CP rate asymmetry turns out to be
negligibly small (of order $10^{-12}$) in $\tau^{\pm} \to K^{\pm}
\pi^0\nu_{\tau}$ decays \cite{Delepine:2005tw}, opening a large window to
consider the effects of New Physics contributions.

Supersymmetry (SUSY) is one of the most interesting candidates for physics
beyond the SM. In SUSY models there are new sources of CP and flavor
violation that may lead to significant impacts on the CP asymmetries of $%
\tau $-decays. In this paper we analyze SUSY contributions to the CP
violating effects in $\tau \to K \pi \nu_{\tau}$ decays. We consider the
effects due to the chargino and neutralino exchanges by using the mass
insertion approximation (MIA) which is a very effective tool for studying
SUSY contributions to flavor changing neutral current processes (FCNC) in a
model independent way. We take into account all the relevant operators
involved in the effective Hamiltonian for $\vert\Delta S\vert = 1$ $\tau$
decays and provide analytical expression for the corresponding Wilson
coefficients.

The elementary process underlying $\vert\Delta S \vert=1$ decays is $\tau^-
\to s \bar{u}\nu_{\tau}$. The lowest order contribution to this decay in the
SM is mediated by the exchange of a single $W^-$ boson, which is Cabibbo
suppressed. We consider in this paper the higher order effects induced by
supersymmetry in the effective Hamiltonian to describe this low-energy
process. The observable effects induced in some of the dominant $|\Delta S|
=1$ exclusive processes are considered. The main focus of our paper is to
provide an specific mechanism to generate CP-violating couplings in the
scalar form factor of $\tau \to K \pi\nu_{\tau}$ decays as studied by the
CLEO collaboration in \cite{Bonvicini:2001xz}.

This paper is organized as follows. In section 2 we briefly review
the SM contribution to the $\vert \Delta S \vert=1$ $\tau $
decays. As pointed out before, the resulting CP rate asymmetry in
$\tau^{\pm} \to K^{\pm}\pi^0 \nu_{\tau}$ is negligible. Moreover,
we show that this result remains intact in the case of minimal
extension of the SM with right handed neutrinos. In section 3 we
derive the SUSY effective hamiltonian for the $\tau^- \to s
\bar{u} \nu_{\tau}$ transitions. In Section 4 we consider the
effects of SUSY contributions on the branching ratios of two
dominant exclusive $\vert \Delta S\vert=1$ decays, namely $\tau^-
\to K^- \nu_{\tau}$ and $\tau^- \to (K\pi)^- \nu_{\tau}$. Section
5 is devoted for analyzing the SUSY contributions to the CP
asymmetry in $\tau \to K \pi \nu_{\tau}$ decay. We show that
within SUSY models, one can generate the CP asymmetry, however it
is below the experimental limits. Finally, we give our main
conclusions in section 6. We have also included an Appendix to
provide the complete expressions of the Wilson coefficients
derived from SUSY.

%%%%%%%%%%%%%%%%%%%%%%%%%%%%%%%%%%%%%%%%%%%%%%%%%%%%%%%%%%%%%%%%%%%%%

\section{{\protect\large \textbf{$\vert \Delta S \vert=1$ $\protect\tau $
decays in the Standard Model}}}

\begin{figure}[h]
\begin{center}
\epsfig{file=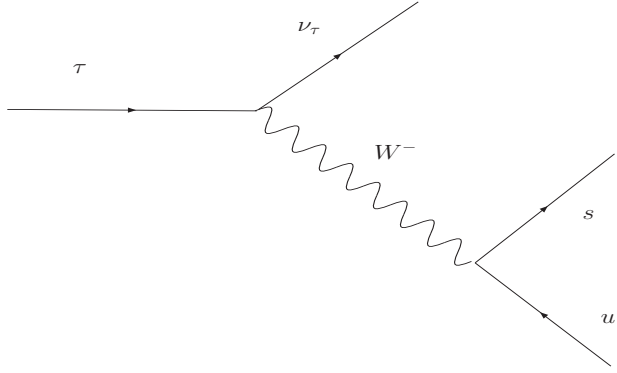, width=9cm, height=5cm, angle=0}
\end{center}
\caption{SM tree-level contributions to $\protect\tau^- \to \bar{u} s \protect\nu_{%
\protect\tau}$ transition.}
\label{SMfig}
\end{figure}

Strangeness-changing $\vert \Delta S \vert=1$ decays of tau leptons are
driven by the $\tau^- \to \bar{u}s\nu_{\tau}$ elementary process. In the SM
they occur at the tree-level, as shown in Fig. \ref{SMfig}.
%----------------------------------------------
%-------------------------------------------------
The SM effective Hamiltonian underlying these decays is given by
\begin{equation}
\mathcal{H}_{SM}= \frac{G_F}{\sqrt{2}}V_{us}(\bar{\nu_{\tau}} \gamma^{\mu} L
\tau)(\bar{s}\gamma_{\mu} L u)\ ,
\end{equation}
where $V_{us}$ is the $u\bar{s}$ Cabibbo-Kobayashi-Maskawa (CKM) matrix
element and $L,R=1\mp \gamma_5$. The amplitudes for the dominant exclusive
processes derived from this Hamiltonian are:
\begin{eqnarray}
\mathcal{A}^{SM} \left(\tau^-(p) \to K^-(q) \nu_{\tau}(p^{\prime})\right)
&=& i\frac{G_F}{\sqrt{2}}V_{us} f_Km_{\tau} r \\
\mathcal{A}^{SM} \Big(\tau^-(p) \to K(q)\pi(q^{\prime})
\nu_{\tau}(p^{\prime})\Big) &=& i\frac{G_F}{\sqrt{2}}V_{us}C_{\scriptscriptstyle K} \Big[%
f_V(t)~Q^{\mu}~\bar{\nu}(p^{\prime})\gamma_{\mu}L
\tau(p)\nonumber\\ &+& m_{\tau}~ f_S(t)~ r\Big]~~
\end{eqnarray}
where letters within parenthesis denote the momenta of the particles, $f_K$
is the $K^-$ decay constant, $t=(q+q^{\prime})^2$ is the square of the
momentum transfer, $C_{\scriptscriptstyle K}=1 (1/\sqrt{2})$ for $K^0\pi^-(K^-\pi^0)$ state, $r=%
\bar{\nu}(p^{\prime})R \tau(p)$, $\Delta^2 = m_K^2 - m_{\pi}^2$,
and
\begin{equation}
Q_{\mu}=(q - q^{\prime})_{\mu}-
\frac{\Delta^2}{t}(q+q^{\prime})_{\mu}\ .
\end{equation}

In the SM, the two-body decay of Eq.(2) is a clean prediction if one uses $%
f_K \simeq 159.8$ MeV from $K^- \to \mu^-\nu_{\mu}$ decay \cite%
{Eidelman:2004wy}. Since New Physics can affect $\tau \to K\nu$ and $K \to
\mu \nu$ decays in a non-universal way, these decays can be used to obtain
interesting bounds on new physics couplings.

On the other hand, the three-body decay of Eq.(3) can exhibit
eventually the effects of CP violation
\cite{Kuhn:1996dv,Bigi:2005ts,Delepine:2005tw,Bonvicini:2001xz}.
However, the decay rate of this process is given by
\cite{GodinaNava:1995jb}%
\be %
\Gamma\left(\tau \to K \pi \nu_{\tau}\right) = \frac{G_F^2
m_{\tau}^5}{768 \pi^3} \vert V_{us} \vert^2 ~ I_{SM}, %
\ee %
where %
\bea %
I_{SM} &=& \frac{1}{m_{\tau}^6}
\int_{(m_K+m_{\pi})^2}^{m_{\tau}^2} \frac{dt}{t^3}~
\left(m_{\tau}^2 -t\right)^2 \Big[\vert f_V \vert^2
\Big(1+\frac{2t}{m_{\tau}^2}\Big)
[\lambda(t,m_K^2,m_{\pi}^2)]^{3/2} \nonumber\\
&+& 3 \vert f_S \vert^2 \Delta^4
[\lambda(t,m_K^2,m_{\pi}^2)]^{1/2}\Big]. \eea %
The function $\lambda(x,y,z)$ is given by $\lambda(x,y,z) =
x^2+y^2 +z^2 -2 xy -2 xz -2 yz$. It is clear from the expression
of $\Gamma(\tau \to K \pi \nu_{\tau})$ that within the SM, the
direct CP asymmetry identically vanishes. As is well known, a
necessary condition to generate a CP rate asymmetry is that at
least two terms of the amplitude for a given physical process have
different weak and strong phases. However, in the SM the relative
weak phase between the scalar $f_S$ and the vector $f_V$ form
factors of $\tau \to K\pi\nu$ is zero. Furthermore, since the form
factors $f_{V,S}(t)$ in Eq. (3) belong respectively to the
(orthogonal) $l=1$ and $l=0$ angular momentum configurations of
the $K\pi$ system, the $f_S f_V$ term in the squared amplitude
vanish upon the integration over the variable
$u=(p-p^{\prime})^2$, therefore the CP-violating terms vanish in
the integrated rate $\Gamma$ and in the hadronic spectrum
$d\Gamma/dt$.  Thus, within the SM CP violating effects in this
three-body channel can manifest only in the double differential
decay distribution where the interference of $f_V$ and $f_S$ is
present (see ref. \cite{Bonvicini:2001xz}).

A different mechanism to generate a CP rate asymmetry in the SM for $%
\tau^{\pm} \to K^{\pm}\pi^0\nu_{\tau}$ was considered in \cite%
{Delepine:2005tw}. In this case the two amplitudes with different
weak and strong phases contribute to the same $l=1$ angular
momentum configuration. This asymmetry turns out to be negligibly
small since it is suppressed by the CKM factor $V_{td} \simeq
10^{-3}$ and also by a higher order suppresion factor $g^2/4\pi
M^2_W \simeq10^{-8}$. Thus, the resulting CP rate asymmetry is
expected to be negligible, as confirmed in
Ref.\cite{Delepine:2005tw}. Therefore, this decay can be suitable
to search for the effects of CP violation induced by New Physics.

The minimal extension of the SM with right handed neutrinos $\nu$SM, where
non vanishing neutrino masses can be obtained, allows for a new source of
the CP violation through the $U_{MNS}$ mixing matrix. In this scenario, the
amplitude of the decay $\tau^- \to K^- \pi^0 \nu_{\tau}$ will be given by
\begin{equation}
\mathcal{A^{\nu SM}}(\tau^- \to K^- \pi^0 \nu_{\tau} ) = (U^*_{MNS})_{33}
\vert \mathcal{A^{SM}} \vert ~ e^{i\delta_{SM}}.
\end{equation}
It is remarkable that, although the amplitude $\mathcal{A^{\nu SM}}$ has a
weak CP violating phase, the CP asymmetry still vanishes. Therefore, any
measurement of a non-vanishing CP asymmetry will be a hint for a new physics
beyond the SM. In the rest of the paper we will focus on the NP
contributions to CP violation in the three-body decay induced by SUSY.

\section{{\protect\large \textbf{SUSY contributions to $\vert \Delta
S\vert=1 $ $\protect\tau$ decays}}}

\begin{figure}[h]
\begin{center}
\epsfig{file=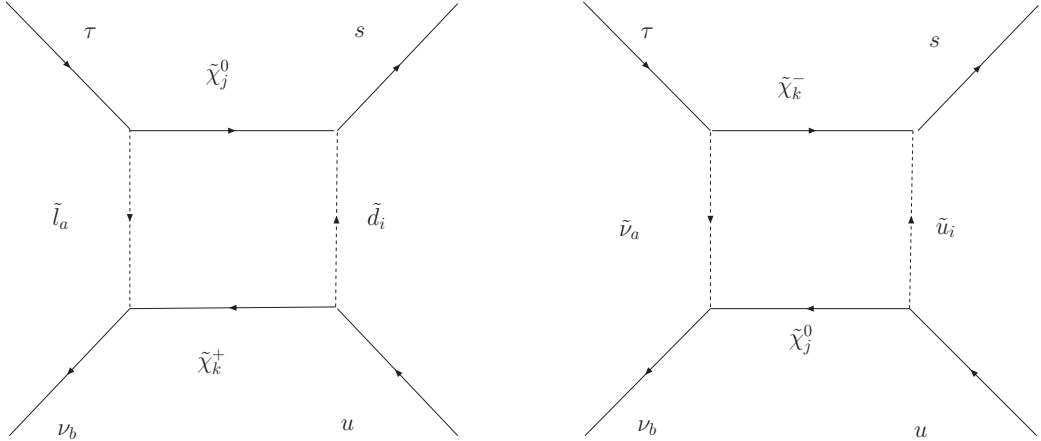, width=14cm, height=6cm,
angle=0}
\end{center}
\caption{SUSY box contributions to $\protect\tau ^{-}\rightarrow \bar{u}s%
\protect\nu _{\protect\tau }$ transition.}
\label{SUSYfig}
\end{figure}

\begin{figure}[h]
\begin{center}
\epsfig{file=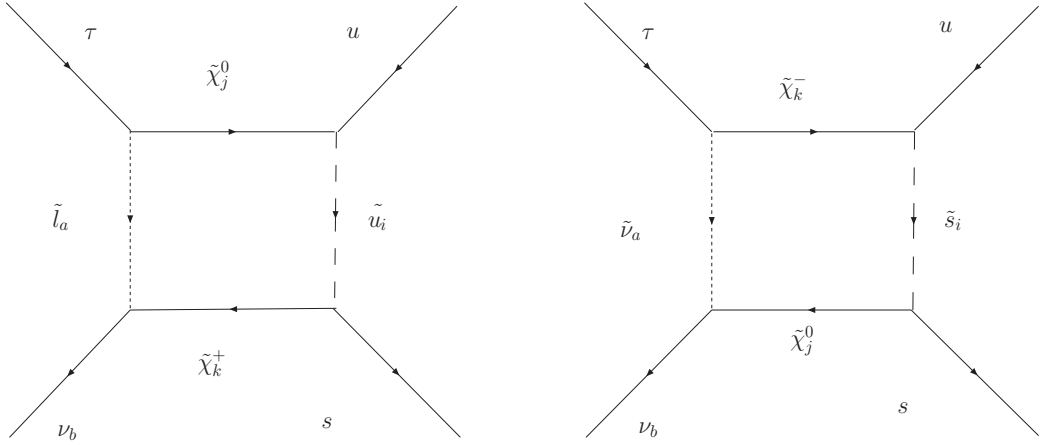, width=14cm, height=6cm,
angle=0}
\end{center}
\caption{Crossed diagrams of Figure 2.}
\label{SUSYfig2}
\end{figure}

The effective Hamiltonian $H_{eff}$ derived from SUSY can be expressed as
\begin{equation}
H_{eff}=\frac{G_{F}}{\sqrt{2}}V_{us}\sum_{i}C_{i}(\mu )Q_{i}(\mu
), \label{SHeff}
\end{equation}%
where $C_{i}$ are the Wilson coefficients and $Q_{i}$ are the relevant local
operators at low energy scale $\mu \simeq m_{\tau }$. The operators are
given by
\begin{eqnarray}
Q_{1} &=&(\bar{\nu}\gamma ^{\mu }L\tau )(\bar{s}\gamma _{\mu }Lu), \\
Q_{2} &=&(\bar{\nu}\gamma _{\mu }L\tau )(\bar{s}\gamma _{\mu }Ru), \\
Q_{3} &=&(\bar{\nu}R\tau )(\bar{s}Lu), \\
Q_{4} &=&(\bar{\nu}R\tau )(\bar{s}Ru), \\
Q_{5} &=&(\bar{\nu}\sigma _{\mu \nu }R\tau )(\bar{s}\sigma ^{\mu \nu }Ru).
\end{eqnarray}%
where $L,R$ are as defined in the previous section and $\sigma ^{\mu \nu }=%
\frac{i}{2}[\gamma ^{\mu },\gamma ^{\nu }]$. The Wilson coefficients $C_{i}$%
, at the electroweak scale, can be expressed as $%
C_{i}=C_{i}^{SM}+C_{i}^{SUSY}$ where $C_{i}^{SM}$ are given by
\begin{eqnarray}
&&C_{1}^{SM}=1,  \nonumber \\
&&C_{2,3,4,5}^{SM}=0.
\end{eqnarray}%
SUSY contributions to the Hamiltonian of $\tau ^{-}\rightarrow \bar{u}s\nu
_{\tau }$ transitions can be generated through two topological box diagrams
as shown in Figs. (\ref{SUSYfig},\ref{SUSYfig2}). Other SUSY contributions
(vertex corrections) are suppressed either due to small Yukawa couplings of
light quarks or because they have the same structure as the SM in the
hadronic vertex. %----------------------------------------------
%------------------------------------------
%------------------------------------------
In our computations of Wilson coefficients, we will work in the mass
insertion approximation (MIA), where gluino and neutralino are flavor
diagonal. Denoting by $(\Delta _{AB}^{f})_{ab}$ the \textit{off-}diagonal
terms in the sfermion mass matrices where $A,B$ indicate chirality, $%
A,B=(L,R)$, the $A-B$ sfermion propagator can be expanded as
\begin{equation}
\langle \tilde{f}_{A}^{a}\tilde{f}_{B}^{b\ast }\rangle =i(k^{2}I-\tilde{m}%
^{2}I-\Delta _{AB}^{f}))_{ab}^{-1}\simeq \frac{i\delta _{ab}}{k^{2}-\tilde{m}%
^{2}}+\frac{i(\Delta _{AB}^{f})_{ab}}{(k^{2}-\tilde{m}^{2})^{2}}+O(\Delta
^{2}),
\end{equation}%
where $\tilde{f}$ denotes any scalar fermion, $a,b=(1,2,3)$ are flavor
indices, $I$ is the unit matrix, and $\tilde{m}$ is the average sfermion
mass. It is convenient to define a dimensionless quantity $(\delta
_{AB}^{f})_{ab}\equiv (\Delta _{AB}^{f})_{ab}/\tilde{m}^{2}.$ As long as $%
(\Delta _{AB}^{f})_{ab}$ is smaller than $\tilde{m}^{2}$ we can consider
only the first order term in $(\delta _{AB}^{f})_{ab}$ of the sfermion
propagator expansion. In our analysis we will keep only terms proportional
to the third generation Yukawa couplings and terms of order $\lambda $ where
$\lambda =V_{us}.$

The complete expressions for the Wilson coefficients $C_i$ at
$m_W$ scale induced by SUSY computed from Figs.
(\ref{SUSYfig},\ref{SUSYfig2}) can be found in the Appendix. As
can be seen from this appendix, the $C_i$ are given in terms of
several mass insertions that represent the flavor transitions
between different generations of quarks or leptons. In general,
these mass insertions are complex and of order one. However, the
experimental limits of several flavor changing neutral currents
impose severe constraints on most of these mass insertions. In the
following, we summarize all the important constraints on the
relevant mass insertions for our process.

\begin{enumerate}
\item From the experimental measurements of $BR(\mu \to e \gamma) < 1.2
\times 10^{-11}$, the following bounds on $\vert(\delta^l_{12})_{AB}\vert$
and $\vert(\delta^{\nu}_{12})_{AB}\vert$ are obtained \cite{Branco:2003zy}:
For $M_1 \sim M_2=100$ GeV and $\mu =\tilde{m}_l=200$ GeV,
\begin{eqnarray}
&&\vert(\delta^l_{12})_{LL}\vert \raise0.3ex\hbox{$\;<$\kern-0.75em%
\raise-1.1ex\hbox{$\sim\;$}} 10^{-3}, ~~~~~~~~
\vert(\delta^l_{12})_{LR}\vert \raise0.3ex\hbox{$\;<$\kern-0.75em%
\raise-1.1ex\hbox{$\sim\;$}} 10^{-6}, \\
&&\vert(\delta^{\nu}_{12})_{LL}\vert \raise0.3ex\hbox{$\;<$\kern-0.75em%
\raise-1.1ex\hbox{$\sim\;$}} 6\times 10^{-4}, ~~~
\vert(\delta^{\nu}_{13})_{LL}\vert \raise0.3ex\hbox{$\;<$\kern-0.75em%
\raise-1.1ex\hbox{$\sim\;$}} 4\times 10^{-4}, ~~~~
\vert(\delta^{\nu}_{23})_{LL}\vert \raise0.3ex\hbox{$\;<$\kern-0.75em%
\raise-1.1ex\hbox{$\sim\;$}} 7\times 10^{-4}.~~~~~~~
\end{eqnarray}

\item From $BR(\tau \to \mu \gamma) < 1.1 \times 10^{-6}$, one gets the
following constraint on $\vert (\delta^l_{23})_{LR})\vert$ \cite%
{Gabbiani:1996hi}:
\begin{equation}
\vert(\delta^l_{23})_{LR}\vert \raise0.3ex\hbox{$\;<$\kern-0.75em%
\raise-1.1ex\hbox{$\sim\;$}} 2 \times 10^{-2},
\end{equation}
and from $BR(\tau \to e \gamma) < 2.7 \times 10^{-6}$, one finds \cite%
{Gabbiani:1996hi}:
\begin{equation}
\vert(\delta^l_{13})_{LR}\vert \raise0.3ex\hbox{$\;<$\kern-0.75em%
\raise-1.1ex\hbox{$\sim\;$}} 1 \times 10^{-1}.
\end{equation}

\item The mass insertions $(\delta_{12}^d)_{AB}$ are constrained by the $%
\Delta M_K $ and $\epsilon_K $ as follows \cite{Gabbiani:1996hi}:
\begin{eqnarray}
\vert(\delta^d_{12})_{LL}\vert \raise0.3ex\hbox{$\;<$\kern-0.75em%
\raise-1.1ex\hbox{$\sim\;$}} 4\times 10^{-2}, ~~~~~~~~~~~~~~~~
\vert(\delta^d_{12})_{LR}\vert \raise0.3ex\hbox{$\;<$\kern-0.75em%
\raise-1.1ex\hbox{$\sim\;$}} 4 \times 10^{-3}, \\
\sqrt{\vert\mathrm{{Im}\left[(\delta^d_{12})_{LL}\right]^2\vert}} \raise0.3ex%
\hbox{$\;<$\kern-0.75em\raise-1.1ex\hbox{$\sim\;$}} 3\times 10^{-3}, ~~~~~~
\sqrt{\vert\mathrm{{Im}\left[(\delta^d_{12})_{LR}\right]^2\vert}} \raise0.3ex%
\hbox{$\;<$\kern-0.75em\raise-1.1ex\hbox{$\sim\;$}} 3\times 10^{-4}.
\end{eqnarray}

\item The mass insertion $(\delta_{12}^u)_{AB}$ are constrained by the $%
\Delta M_D$ as follows \cite{Khalil:2006zb}:
\begin{equation}
\vert(\delta^u_{12})_{LL}\vert \raise0.3ex\hbox{$\;<$\kern-0.75em%
\raise-1.1ex\hbox{$\sim\;$}} 1.7\times 10^{-2}, ~~~~~~~~~~~~~~~
\vert(\delta^u_{12})_{LR}\vert \raise0.3ex\hbox{$\;<$\kern-0.75em%
\raise-1.1ex\hbox{$\sim\;$}} 2.4 \times 10^{-2}.
\end{equation}
\end{enumerate}

Here, three comments are in order. $i)$ Due to the Hermiticity of the $LL$
sector in the sfermion mass matrix, $(\delta _{AB}^{f})_{LL}=(\delta
_{AB}^{f})_{LL}^{\dagger }=(\delta _{BA}^{f})_{LL}^{\ast }$, where $%
A,B=1,2,3 $. $ii)$ The above constraints imposed on the mass insertions $%
(\delta _{AB}^{q,l})_{LL,LR}$ are derived from supersymmetric
contributions through exchange of gluino or neutralino which
preserves chirality, therefore same constraints are also imposed
on the mass insertions $(\delta _{AB}^{q,l})_{RR,RL}$. $iii)$ The
mass insertions $(\delta
_{AB}^{f})_{LR(RL)}$ are not, in general, related to the mass insertions $%
(\delta _{BA}^{f})_{LR(RL)}$. Taking the above constraints into account, one
finds that the dominant contribution to the $\tau ^{-}\rightarrow u\bar{s}%
\nu _{\tau }$ is given in terms of $(\delta _{32}^{\nu })_{LR}$, $(\delta
_{32}^{\nu })_{RL}$, $(\delta _{21}^{d})_{RL}$ and $(\delta _{21}^{u})_{LR}$%
. Notice that the effective Hamiltonian (eq. \ref{SHeff}) derived
in this section can induce supersymmetric effects in all the
$|\Delta S|=1$ exclusive $\tau $ lepton decay. In the following
section we consider two examples.

\section{\large{\bf Constraints from two-body $\tau$-decays}}

In this section we analyze the possible constraints that may be
imposed on the SUSY contributions to $\tau \to \bar{s}~ u~
\nu_{\tau}$ from the exclusive $|\Delta S|=1$ decay: $\tau
^{-}(p)\rightarrow K^{-}(q)~\nu _{\tau }(p^{\prime })$. The decay
amplitude considering effects of SUSY contributions reads:
\begin{equation}
\mathcal{A}(\tau ^{-}\rightarrow K^{-}\nu _{\tau })=\mathcal{A}_{SM}+%
\mathcal{A}_{SUSY}\ ,
\end{equation}%
which can be written explicitly as:
\begin{equation}
\mathcal{A}(\tau ^{-}\rightarrow K^{-}\nu _{\tau })=i\frac{G_{F}}{\sqrt{2}}%
V_{us}f_{K}~m_{\tau }~(\bar{\nu}(p^{\prime }) R \tau(p))\left\{
1+\delta _{SUSY}(\tau )\right\} \ .
\end{equation}%
The decay $K^{-}\rightarrow \mu ^{-}\nu $, which fixes $f_{K}$ in the
absence of new physics, would also be modified by the effects of new
physics:
\begin{equation}
\mathcal{A}(K^{-}\rightarrow \mu ^{-}\nu )=i\frac{G_{F}}{\sqrt{2}}%
V_{us}f_{K} m_{\mu}~ (\bar{\nu}(p^{\prime }) R \mu(p))\left\{
1+\delta _{SUSY}(K)\right\} \ .
\end{equation}%
In order to estimate the size of SUSY contributions in such decays, one
defines the ratio:
\begin{eqnarray*}
R_{\tau /K} &=&\frac{\Gamma (\tau \rightarrow K\nu {\tau }(\gamma ))}{\Gamma
(K\rightarrow \mu \nu (\gamma ))} \\
&=&\frac{m_{\tau }^{3}}{2m_{K}m_{\mu }^{2}}.\frac{\left( 1-\frac{\textstyle %
m_{K}^{2}}{\textstyle m_{\tau }^{2}}\right) ^{2}}{\left( 1-\frac{\textstyle %
m_{\mu }^{2}}{\textstyle m_{K}^{2}}\right) ^{2}}.(1+\delta R_{\tau /K})\left[
1+2Re(\delta _{SUSY}(\tau )-\delta _{SUSY}(K))\right] \ .
\end{eqnarray*}%
In the above equation $(\gamma )$ means that complete SM radiative
corrections of $O(\alpha )$ have been included. The long-distance radiative
corrections, which do not cancel in the above ratio, were computed in \cite%
{Decker:1994ea} and reads $\delta R_{\tau /K}=(0.90_{-0.26}^{+0.17})\%$ .
Using the experimental rates for the involved decays: $\Gamma (\tau
\rightarrow K\nu )=(2.36\pm 0.08)\times 10^{10}s^{-1},\Gamma (K\rightarrow
\mu \nu )=(0.5118\pm 0.0018)\times 10^{8}s^{-1}$ \cite{Eidelman:2004wy} into
eq. (21) we get:
\begin{equation}
Re[\delta _{SUSY}(K)-\delta _{SUSY}(\tau )]=0.02\pm 0.03.  \label{exp1}
\end{equation}%
This equation is actually a model-independent result. Using the
expression for the Wilson coefficient induced by SUSY, it is easy
to see that the dominant contribution to $Re(\delta _{SUSY}(\tau
)-\delta _{SUSY}(K))$ comes from $C_{1}^{SUSY}$. In terms of the
SUSY parameters, it can be translated as follows
\begin{equation}
Re[\delta _{SUSY}(K)-\delta _{SUSY}(\tau )]\simeq
Re[C^{SUSY}_{1}(K) - C^{SUSY}_{1}(\tau )]
\end{equation}%
where $C^{SUSY}_{1}(K,\tau )$ are respectively the Wilson
coefficients corresponding to the operators responsible of
$K\rightarrow \mu \nu _{\mu }$ and $\tau \rightarrow K\pi \nu
_{\tau }$. Using as input parameters $M_{1}=100$, $M_{2}=200$
GeV  and $\mu =M_{\tilde{q}}= 400$ GeV and $\tan \beta \simeq 20$, one gets%
\be Re[C^{SUSY}_{1}(K)-C^{SUSY}_{1}(\tau )]\simeq -0.03(\delta
_{21}^{u})_{LL}-
0.006(\delta _{23}^{l})_{LL}%
\ee%
where we keep the dominant contributions only. Using the bounds on the $\delta $%
's given in previous section, the terms which is mostly unconstraint is $%
(\delta _{23}^{l})_{LL}$. It is clear that still we are far from the
experimental limit given in (\ref{exp1}) but it is important to notice that
experimental data should strongly improve in a close future\cite%
{Igonkina:2006tv} . %
%%%%%%%%%%%%%%%%%%%%%%%%%%%%%%%%%%%%%%%%%%%%%%%%%%%%
\section{\large{\bf SUSY contribution to CP asymmetry in $ \tau \to K \pi \nu_{\tau}$}}

Having analyzed the constraints from two-body $\tau$-decays
imposed on the supersymmetric contributions to $\tau \to \bar{s}~
u~ \nu_{\tau}$ transition, now we can study the supersymmetric
effects on the CP violation in the three-body decay $ \tau \to K
\pi \nu_{\tau}$. We will show that although supersymmetry enhances
the asymmetry of this process by many order of magnitude than the
SM expectation, the resulting CP asymmetries are still smaller
than the current experimental reaches.

Given the spin-parity properties of the $K \pi $ system, we can
write the total amplitude (SM and SUSY) of the $\tau(p)\rightarrow
K(q)\pi (q^{\prime })\nu _{\tau
}(p^{\prime })$ decay as%
\bea %
\mathcal{A}_{T}(\tau \rightarrow K\pi \nu ) &=&\frac{G_{F}V_{us}}{\sqrt{2}}%
\Big[ (1+C_{1})\langle K\pi |\bar{s}\gamma _{\mu }u|0\rangle \bar{\nu}%
(p^{\prime })\gamma ^{\mu }L \tau(p) \\
&+&(C_{3}+C_{4})\langle K\pi |\bar{s}u|0\rangle \bar{\nu}%
(p^{\prime })R \tau(p) + C_{5}\langle K\pi |\bar{s}\sigma _{\mu
\upsilon }u|0\rangle
\bar{\nu}(p^{\prime })\sigma ^{\mu \upsilon }R \tau(p)\Big] , \nonumber %
\eea %
where $C_i$ stand for $C_i^{SUSY}$, since the $C_i^{SM}$ are
explicitly included . It is now clear that the resulting CP
asymmetry depends on the relative ratio among the SUSY Wilson
coefficients. For example, in case that $C_1$ is giving the
dominant SUSY contribution and $C_{3,4,5}$ effects can be
negligible, then the CP asymmetry of $\tau \to K \pi \nu_{\tau}$
will vanish identically as in the SM.
We consider the following two interesting scenarios:\\

$i)$ The case of $C_{3}$ or $C_{4}$ gives relevant contributions
while $C_5$ is negligible. In this case, SUSY induces a relative
weak phase between the vector and scalar form factors describing
this process.\\

$ii)$ The case of $C_{5}$ gives relevant contributions while
$C_{3,4}$ are negligible. In this case, SUSY induces a relative
weak phase between the vector and tensor form factors.\\

Let us start by analyzing the CP asymmetry in the first scenario.
Using the definition of the hadronic matrix element introduced in
Eq.(3):
\begin{equation}
\langle K\pi |\bar{s}\gamma _{\mu }u|0\rangle
=C_{\scriptscriptstyle K}\{f_{V}(t)Q_{\mu }+f_{S}(t)(q+q^{\prime
})_{\mu }\}~,
\end{equation}%
we can obtain the hadronic matrix element of the scalar current by taking
the divergence in the usual form:
\begin{equation}
\langle K\pi |\bar{s}u|0\rangle =\frac{C_{K}t}{m_{s}-m_{u}}f_{S}(t)~,
\end{equation}%
where $m_{s,u}$ denote $s,u$ current quark masses. Thus, we finally get the
amplitude:
\begin{eqnarray}
\mathcal{A}_{T}(\tau \rightarrow K\pi \nu ) &=&\frac{C_{K}G_{F}V_{us}}{\sqrt{%
2}}(1+C_{1})\times   \nonumber \\
&&\!\!\!\!\!\!\!\!\left\{ f_{V}Q^{\mu }\bar{u}(p^{\prime })\gamma _{\mu
}Lu(p)+\left[ m_{\tau }+\left( \frac{C_{3}+C_{4}}{1+C_{1}}\right) \frac{t}{%
m_{s}-m_{u}}\right] f_{S}\bar{u}(p^{\prime })Ru(p)\right\} \
.\nonumber
\end{eqnarray}%
It is remarkable that in this case, SUSY effects just modify the
normalization of the SM amplitude and the relative size of the
vector and scalar contributions.

When we compare this expression with the decay amplitude given in
Eq. (2) of Ref. \cite{Bonvicini:2001xz}:
\begin{equation}
\mathcal{A}(\tau^-\to K \pi\nu_{\tau}) \sim {\bar{u}(p^{\prime})}%
\gamma_{\mu}Lu(p)f_V Q^{\mu} +\Lambda {\bar{u}(p^{\prime})}Ru(p)f_SM\ ,
\end{equation}
where $M=1$ GeV is a normalization mass scale, we obtain the relation
\begin{equation}
\Lambda M=m_{\tau}+
\left(\frac{C_3+C_4}{1+C_1}\right)\frac{t}{m_s-m_u}\ .
\label{cleo-Lambda}
\end{equation}
The first term in the r.h.s of this equation is the usual contribution of
the SM, which is real, and the second term arises from the SUSY
contributions and contains a CP-violating phase, hence it can generate a
non-vanishing CP asymmetry.

The square of the matrix element becomes:
\begin{eqnarray}
\sum_{pols}\vert\mathcal{A}\vert^2&\sim &\vert
f_V\vert^2(2p.Qp^{\prime}.Q-p.p^{\prime}Q^2)+\vert\Lambda \vert^2 \vert
f_S(t)\vert^2M^2p.p^{\prime}  \nonumber \\
&& +2Re\Lambda\cdot Re(f_Sf_V^*)M m_{\tau}p^{\prime}.Q -2Im\Lambda\cdot
Im(f_Sf_V^*)Mm_{\tau}p^{\prime}.Q
\end{eqnarray}

The last term in the previous equation is odd under a CP
transformation but we should notice that the last two terms
disappear once we integrate on the kinematical variable $u$ of the
process in consideration. This means that it is not possible to
generate a CP asymmetry in total decay rates corresponding to this
process. So the only way to generate a CP asymmetry is to look for
the double differential distribution $(d^{2}\Gamma /dudt)$ or a
variant of it as CLEO collaboration did in
Ref.\cite{Bonvicini:2001xz}. The CLEO collaboration has recently
studied the ratio of CP-odd to CP-even terms of this squared
amplitude for $\tau ^{\pm }\rightarrow K_{S}\pi ^{\pm }\nu _{\tau
}$ decays and has obtained the following bound: $-0.172<Im(\Lambda
)<0.067$ at 90\% C.L. Using Eq.(\ref{cleo-Lambda}), we can
translate this bound into:
\begin{equation}
-0.010\leq Im\left( \frac{C_{3}+C_{4}}{1+C_{1}}\right) \leq 0.004\ ,
\end{equation}%
where we have used $m_{s}-m_{u}=100$ MeV, and the average value
$\langle t\rangle \approx (1332.8\ \mbox{\rm MeV})^{2}$. Now, as
input parameters $M_{1}=100$ and $M_{2}=200$ GeV \ and $\mu
=M_{\tilde{q}}=400$ GeV and $\tan \beta =20$, one gets
\begin{equation}
Im\left( \frac{C_{3}+C_{4}}{1+C_{1}}\right) \simeq 1.3\times
10^{-5}Im(\delta _{21}^{d})_{RL}
\end{equation}
Still the experimental bound is too loose to give us information on $%
(\delta _{21}^{d})_{RL}$. Notice however that forthcoming measurements of
the CP asymmetry in the $K_{S}\pi $ channel will be significantly improved
at B factories \cite{Igonkina:2006tv} since their data sample is larger by
two order of magnitude than the one used by CLEO \cite{Bonvicini:2001xz} in
their analysis.

Now, we turn to take into consideration the $O_{5}$ operator which
is naturally induced by SUSY corrections to Wilson coefficients.
This operator could interfere with $O_{1}$ operator which contains
SM contributions and the strong phases. So in principle, using
this interference between $O_{5}$ and $O_{1}$, it should be
possible to generate a CP asymmetry directly in the total decay
rate which is completely forbidden in SM. Let us keep the dominant
contribution to $\tau ^{-}\rightarrow (K\pi )^{-}\nu _{\tau }$
and the $O_{5}$ contribution:%
\begin{eqnarray}
\mathcal{A}_{T}(\tau \rightarrow K\pi \nu ) &=& \frac{G_{F}V_{us}}{\sqrt{2}}(1+C_{1})
\left\{ f_{V}(t)Q_{\mu }\bar{u}%
(p^{\prime })\gamma ^{\mu }Lu(p)\right.  \nonumber \\
&+&\left. \frac{C_{5}}{1+C_{1}}\langle K\pi |\bar{s}\sigma _{\mu
\upsilon }u|0\rangle \bar{u}(p^{\prime })\sigma ^{\mu \upsilon
}Ru(p)\right\}
\end{eqnarray}%
In this expression, we have neglected $f_{S}$ effect since its
contribution to total decay rate is numerically small (around
$3\%$ at most, see\cite{GodinaNava:1995jb}) and conserves CP. The
most general form of the antisymmetric matrix element of the
hadronic tensor
current is given by%
\begin{equation}
\langle K\pi |\bar{s}\sigma _{\mu \upsilon }u|0\rangle =\frac{ia}{m_{K}}%
\left[ (p_{\pi })^{\mu }(p_{K})^{\nu }-(p_{\pi })^{\nu }(p_{K})^{\mu }\right]
\end{equation}%
where $a$ is a dimensionless quantity which fixes the scale of the
hadronic matrix element. It is important to remember that
$f_{V}(t)$
contains the strong phases as it can be parametrized\cite{GodinaNava:1995jb}:%
\begin{equation}
f_{V}(t)=\frac{f_{V}(0)m_{K^{\ast }}^{2}}{m_{K^{\ast }}^{2}-t-im_{K^{\ast
}}\Gamma _{K^{\ast }}}
\end{equation}%
The tensor form factor which is given by%
\begin{equation}
f_{T}=\frac{aC_{5}}{1+C_{1}}
\end{equation}%
has no strong phases but of course could have a weak phase (arises
either from $C_1$ or $C_5$) and it can be at most a slightly
varying function of $t$. So, one can compute now the CP
asymmetry in total decay rate:%
\begin{eqnarray}
a_{CP} &=&\frac{\Gamma (\tau ^{-}\rightarrow K^{-}\pi ^{0}\nu _{\tau
})-\Gamma (\tau ^{+}\rightarrow K^{+}\pi ^{0}\nu _{\tau })}{\Gamma (\tau
^{-}\rightarrow K^{-}\pi ^{0}\nu _{\tau })+\Gamma (\tau ^{+}\rightarrow
K^{+}\pi ^{0}\nu _{\tau })} \\
&=&\frac{a}{\Gamma _{SM}}Im C_{5}\frac{-G_{F}^{2}\left\vert
V_{us}\right\vert ^{2}}{128\pi ^{3}m_{\tau }^{2}m_{K}}\int_{(m_{K}+m_{\pi
})^{2}}^{m_{\tau }^{2}}dt\frac{(m_{\tau }^{2}-t)}{t^{2}} Im \left(
f_{V}(t)\right) \lambda ^{1/2}  \nonumber \\
&&\times \left\{ m_{\pi }^{2}(t+\Delta ^{2})^{2}+m_{\tau }^{2}\lambda +\left[
(t-m_{\pi }^{2})^{2}-m_{K}^{2}(t+m_{\pi }^{2})\right] (t-\Delta ^{2})\right\}
\end{eqnarray}%
where $\lambda$, $\Delta$ and $\Gamma_{SM}$ are given in
section 2. Integrating numerically on $t$, one gets%
\begin{eqnarray}
a_{CP} &\simeq &\frac{a}{2}Im \ C_{5} \\
&\simeq &1.4\times 10^{-7}a \ Im (\delta _{21}^{u})_{LR}
\end{eqnarray}%
where to get the last equation, we use the same SUSY parameters as before.
Again within SUSY extensions of the SM, this CP asymmetry is small (even if
it is practically five orders of magnitude bigger than the one expected in SM%
\cite{Delepine:2005tw}) .\ Clearly the observation of a CP asymmetry in this
channel at a range bigger than $10^{-6}$ will be not only a clear evidence
of Physics beyond Standard Model but also an evidence we need Physics beyond
Supersymmetric extensions of Standard Model.

\section{Conclusion}

In summary, in this paper we have computed the effective
hamiltonian derived from SUSY for $|\Delta S|=1$ tau lepton decays
using the mass insertion approximation. Although experimental data
for such decays are not precise enough at present to give
constraints on the fundamental parameters of SUSY, we have shown
how physics beyond standard model as supersymmetrix extensions of
the SM could induce CP violating asymmetry in the double
differential distribution as CLEO\ collaboration did in
ref.\cite{Bonvicini:2001xz} and could also induce CP asymmetry in
total decay rate due to interference between $O_{5}$ and $O_{1}$
operators. We have argued that any CP asymmetry in the channel
under consideration bigger than $10^{-6}$ will be a clear evidence
of not only Physics beyond Standard Model but also an evidence of
Physics beyond SUSY extensions of the SM. We also provided
model-independent constraint on New Physics contribution to $\tau
\rightarrow K \nu$. In particular, it is interesting to observe
that SUSY can provide a specific mechanism to generate a
CP-violating term in the probability distribution of $\tau
\rightarrow K\pi \nu _{\tau }$ decays. Forthcoming and more
precise data for observables of the exclusive processes considered
in this paper will either provide better constraints on SUSY
parameters or give a mechanism to explain discrepancies with the
SM if they are observed.

\bigskip
\section{Acknowledgements}

\noindent The work of D.D. is supported by Conacyt (Mexico) under
the project SEP-2004-C01-46195 and by PROMEP under project
PROMEP/103.5/06/0918. D.D. wants to thank Cinvestav for their
hospitality. G.F and S.K. would like to thank ICTP for the
hospitality, where part of this work took place. The work of G.L.C
was partially supported by Conacyt.
\bigskip

\section{Appendix}

\appendix
%%%%%%%%%%%%%%%%%%%%%%%%%%%%%%%%%%%%%%%%%%%%%%%%%%%%%%%%%%

\section*{\textbf{{\protect\large {SUSY contributions to Wilson coefficients
of $\protect\tau^- \to s \bar{u} \protect\nu_{\protect\tau}$}}}}

Here we provide the complete expressions for the supersymmeric
contributions, at leading order in MIA, for the Wilson
coefficients of $\tau
^{-}\rightarrow s\bar{u}\nu _{\tau }$ transition, $C_{i}(M_{W})$, $i=1,..,5$%
. As mentioned in section 3, the dominant SUSY contributions are given by
chargino-neutralino box diagram exchanges, as illustrated in Fig. 2.

The effective Hamiltonian $H_{eff}$ derived from SUSY can be expressed as
\begin{eqnarray}
H_{eff} &=&\frac{G_{F}}{\sqrt{2}}V_{us}\sum_{i}C_{i}(\mu )Q_{i}(\mu ), \\
&=&\sum_{i}\tilde{C}_{i}(\mu )Q_{i}(\mu ),
\end{eqnarray}%
where $C_{i}$ are the dimensionless Wilson coefficients and
$Q_{i}$ are the relevant local operators at low energy scale $\mu
\simeq m_{\tau }$. The operators are given by
\begin{eqnarray}
Q_{1} &=&(\bar{\nu}\gamma ^{\mu }L\tau )(\bar{s}\gamma _{\mu }Lu), \\
Q_{2} &=&(\bar{\nu}\gamma _{\mu }L\tau )(\bar{s}\gamma _{\mu }Ru), \\
Q_{3} &=&(\bar{\nu}R\tau )(\bar{s}Lu), \\
Q_{4} &=&(\bar{\nu}R\tau )(\bar{s}Ru), \\
Q_{5} &=&(\bar{\nu}\sigma _{\mu \nu }R\tau )(\bar{s}\sigma ^{\mu \nu }Ru).
\end{eqnarray}

In terms of the vertex, one can write the complete vertex as a product of
the vertex coming from leptonic sector and of the vertex coming from
hadronic sector. In this respect we can also write the Wilson coefficients
as
\begin{eqnarray*}
\tilde{C}_{i} &=&C_{i(\tau -\chi ^{-})}^{l}\left( C_{i(\tau -\chi
^{-}-s)}^{q}+C_{i((\tau -\chi ^{-}-u)}^{q}\right) \\
&&+C_{i(\tau -\chi ^{0})}^{l}\left( C_{i(\tau -\chi ^{0}-s)}^{q}+C_{i((\tau
-\chi ^{0}-u)}^{q}\right)
\end{eqnarray*}%
where the $C_{i}^{l}$ is due to the leptonic vertex and $C_{i}^{q}$ is from
the quark sectors. If we expand $C_{i}^{l,q}$ in terms of the mass
insertions, one finds that the leading contributions are given by
\begin{eqnarray}
\tilde{C}_{1} &=&C_{1(\tau -\chi ^{-})}^{l(0)}\left( C_{1(\tau
-\chi ^{-}-s)}^{q(0)}\tilde{I}(x_{i},x_{j})+ C_{1((\tau -\chi
^{-}-u)}^{q(0)}I(x_{i},x_{j})\right)\nonumber\\
&+& C_{1(\tau -\chi ^{0})}^{l(0)}\left( C_{1(\tau -\chi
^{0}-s)}^{q(0)}I(x_{i},x_{j})+C_{1(\tau -\chi
^{0}-u)}^{q(0)}\tilde{I}(x_{i},x_{j})\right) \nonumber\\
&+&C_{1(\tau -\chi ^{-})}^{l(0)}\left( C_{1(\tau -\chi
^{-}-s)}^{q(1)}\tilde{I}_{n}(x_{i},x_{j})+C_{1(\tau -\chi
^{-}-u)}^{q(1)}I_{n}(x_{i},x_{j})\right) \nonumber\\
&+&C_{1(\tau -\chi ^{0})}^{l(0)}\left( C_{1(\tau -\chi
^{0}-s)}^{q(1)}I_{n}(x_{i},x_{j})+C_{1(\tau -\chi
^{0}-u)}^{q(1)}\tilde{I}_{n}(x_{i},x_{j}\right) \nonumber\\
&+&C_{1(\tau -\chi ^{-})}^{l(1)}\left( C_{1(\tau -\chi
^{-}-s)}^{q(0)}\tilde{I}_{n}(x_{i},x_{j})+C_{1(\tau -\chi
^{-}-u)}^{q(0)}I_{n}(x_{i},x_{j})\right)\nonumber\\
&+& C_{1(\tau -\chi ^{0})}^{l(1)}\left( C_{1(\tau -\chi
^{0}-s)}^{q(0)}I_{n}(x_{i},x_{j})+C_{1(\tau -\chi
^{0}-u)}^{q(0)}\tilde{I}_{n}(x_{i},x_{j})\right) \\
&+&O(\delta ^{2}). \nonumber
\end{eqnarray}
With
\begin{eqnarray}
C_{1(\tau -\chi ^{0})}^{l(0)} &=&\frac{-g^{2}}{\sqrt{2}}\left(
N_{i2}^{\ast }+\tan \theta _{w}N_{i1}^{\ast }\right) U_{j1}^{\ast
}(U_{MNS}^{\ast
})_{33}-y_{\tau }^{2}N_{i3}^{\ast }U_{j2}^{\ast }(U_{MNS}^{\ast })_{33},~~ \\
C_{1(\tau -\chi ^{0}-u)}^{q(0)} &=&(\frac{%
1}{8})\frac{g^{2}}{\sqrt{2}}(N_{i2}^{\ast }+%
\frac{1}{3}\tan \theta _{w}N_{i1}^{\ast })(V_{CKM}^{\ast })_{12}V_{j1}, \\
C_{1(\tau -\chi ^{0}-s)}^{q(0)} &=&(\frac{1}{16})\frac{-g^{2}}{\sqrt{2}}\left( N_{i2}-%
\frac{1}{3}\tan \theta _{w}N_{i1}\right) U_{j1}^{\ast
}(V_{CKM}^{\ast
})_{12} ,\\
C_{1(\tau -\chi ^{0})}^{l(1)} &=&-\frac{g^{2}}{\sqrt{2}}\left(
N_{i2}^{\ast }+\tan \theta _{w}N_{i1}^{\ast }\right) U_{j1}^{\ast
}(U_{MNS}^{\ast
})_{a3}(\delta _{LL}^{l})_{a3} \\
&&+\frac{g}{\sqrt{2}}\left( N_{i2}^{\ast }+\tan \theta
_{w}N_{i1}^{\ast }\right) U_{j2}^{\ast }(U_{MNS}^{\ast
})_{33}(h_{e})_{33}(\delta
_{RL}^{l})_{33} \nonumber\\
&&+g(h_{e})_{33}N_{i3}^{\ast }U_{j1}^{\ast }(\delta
_{LR}^{l})_{a3}(U_{MNS}^{\ast })_{a3} \nonumber\\
&&-(h_{e})_{33}^{2}N_{i3}^{\ast }U_{j2}^{\ast }(\delta
_{RR}^{l})_{33}(U_{MNS}^{\ast })_{33}, \nonumber
\end{eqnarray}
and
\begin{eqnarray}
C_{1(\tau -\chi ^{0}-s)}^{q(1)} &=&(\frac{1}{16})\left(
\frac{-g^{2}}{\sqrt{2}}\left( N_{i2}-\frac{1}{3}\tan \theta
_{w}N_{i1}\right) U_{j1}^{\ast }(V_{CKM}^{\ast
})_{1a}(\delta _{LL}^{d})_{2a}\right. \nonumber\\
&&\left. +\frac{g}{\sqrt{2}}\left( N_{i2}-\frac{1}{3}\tan \theta
_{w}N_{i1}\right) U_{j2}^{\ast }(V_{CKM}^{\ast
})_{13}(h_{d})_{33}(\delta _{LR}^{d})_{23}\right),
\end{eqnarray}
\begin{eqnarray}
C_{1(\tau -\chi^{0}-u)}^{q(1)} &=&(\frac{1}{8})\left(
\frac{g^{2}}{\sqrt{2}} (N_{i2}^{\ast}+\frac{1}{3}\tan
\theta_{W}N_{i1}^{\ast
})(V_{CKM}^{\ast})_{a2}(\delta _{LL}^{u})_{a1}V_{j1} \right.\nonumber\\
&&\left. -\frac{g}{\sqrt{2}}(N_{i2}^{\ast }+\frac{1}{3}\tan
\theta_{W}N_{i1}^{\ast})(h_{u})_{33}V_{j2}(V_{CKM}^{\ast}
)_{32}(\delta_{RL}^{u})_{31}\right).
\end{eqnarray}

\bigskip

\begin{eqnarray}
C_{1(\tau -\chi ^{-})}^{l(0)} &=&\frac{g^{2}}{\sqrt{2}}V_{j1}^{\ast
}(U_{MNS}^{\ast })_{33}(N_{i2}-\tan \theta _{w}N_{i1})-(h_{\nu
})_{33}^{2}V_{j2}^{\ast }(U_{MNS}^{\ast })_{33}N_{i4},~~~~~ \\
C_{1(\tau -\chi ^{-}-u)}^{q(0)} &=&(\frac{1}{16})\left( \frac{g^{2}}{\sqrt{2}}%
V_{j1}(N_{i2}^{\ast }+\frac{1}{3}\tan \theta _{w}N_{i1}^{\ast
})(V_{CKM}^{\ast })_{12}\right) , \\
C_{1((\tau -\chi ^{-}-s)}^{q(0)} &=&(\frac{1}{8})\left( -\frac{g^{2}}{\sqrt{2}}(N_{i2}-%
\frac{1}{3}\tan \theta _{w}N_{i1})(V_{CKM}^{\ast
})_{12}U_{j1}^{\ast }\right),
\end{eqnarray}%
\begin{eqnarray}
C_{1(\tau -\chi ^{-})}^{l(1)} &=&\frac{g^{2}}{\sqrt{2}}V_{j1}^{\ast
}(U_{MNS}^{\ast })_{3a}(N_{i2}-\tan \theta _{w}N_{i1})(\delta
_{LL}^{\nu
})_{3a} \nonumber\\
&&+gV_{j1}^{\ast }(h_{\nu })_{33}N_{i4}(\delta _{RL}^{\nu
})_{3a}(U_{MNS}^{\ast })_{3a} \nonumber\\
&&-\frac{g}{\sqrt{2}}(N_{i2}-\tan \theta _{w}N_{i1})V_{j2}^{\ast
}(h_{\nu
})_{aa}(U_{MNS}^{\ast })_{3a}(\delta _{LR}^{\nu })_{3a}\nonumber \\
&&-(h_{\nu })_{33}N_{i4}V_{j2}^{\ast }(h_{\nu })_{aa}(\delta
_{RR}^{\nu })_{3a}(U_{MNS}^{\ast })_{3a}.
\end{eqnarray}%
In case of decoupling of the sneutrino-right, the last terms are
strongly suppressed.

\begin{eqnarray}
C_{1(\tau -\chi ^{-}-u)}^{q(1)} &=&(\frac{1}{16})\left( \frac{g^{2}}{\sqrt{2}}%
V_{j1}(N_{i2}^{\ast }+\frac{1}{3}\tan \theta _{w}N_{i1}^{\ast
})(V_{CKM}^{\ast })_{a2}(\delta _{LL}^{u})_{a1}\right. \nonumber\\
&&\left. -\frac{g}{\sqrt{2}}(h_{u})_{33}V_{j2}(V_{CKM}^{\ast
})_{32}(\delta _{RL}^{u})_{31}(N_{i2}^{\ast }+\frac{1}{3}\tan \theta
_{w}N_{i1}^{\ast })\right)
\end{eqnarray}%
\begin{eqnarray}
C_{1(\tau -\chi ^{-}-s)}^{q(1)} &=&(\frac{1}{8})\left( -\frac{g^{2}}{\sqrt{2}}(N_{i2}-%
\frac{1}{3}\tan \theta _{w}N_{i1})(V_{CKM}^{\ast })_{1a}U_{j1}^{\ast
}(\delta _{LL}^{d})_{2a}\right. \nonumber\\
&&\left. +\frac{g}{\sqrt{2}}(N_{i2}-\frac{1}{3}\tan \theta
_{w}N_{i1})(V_{CKM}^{\ast })_{13}U_{j2}^{\ast }(h_{d})_{33}(\delta
_{LR}^{d})_{23}\right).
\end{eqnarray}

The contribution to $C_2$ is found to vanish identically,
\textit{i.e.},

\be \tilde{C}_{2}=0, \ee%
\begin{eqnarray}
\tilde{C}_{3} &=&C_{3(\tau -\chi ^{-})}^{l(0)}C_{3(\tau -\chi
^{-}-s)}^{q(1)}I_{n}(x_{i},x_{j})+C_{3(\tau -\chi
^{0})}^{l(0)}C_{3(\tau -\chi ^{0}-s)}^{q(1)}I_{n}(x_{i},x_{j})
\nonumber\\
&&+O(\delta ^{2}),
\end{eqnarray}
where $I_n(x_i,x_j)$ is defined below and
$x_i=m_{\chi_i^{\pm}}^2/\tilde{m}^2$ and
$x_j=m_{\chi_j^{0}}^2/\tilde{m}^2$.
\begin{eqnarray}
C_{3(\tau -\chi ^{0})}^{l(0)} &=&g(h_{e})_{33}N_{i3}U_{j1}^{\ast
}(U_{MNS}^{\ast })_{33} \\
&&-g\sqrt{2}\tan \theta _{w}N_{i1}U_{j2}^{\ast
}(h_{e})_{33}(U_{MNS}^{\ast })_{33},
\end{eqnarray}%
\be C_{3(\tau -\chi
^{-})}^{l(0)}=-(h_{e})_{33}\frac{g}{\sqrt{2}}(N_{i2}-\tan \theta
_{w}N_{i1})U_{j2}(U_{MNS}^{\ast })_{33}, \ee%

\begin{eqnarray}
C_{3(\tau -\chi ^{-}-s)}^{q(1)} &=&(-\frac{1}{8})\left( \frac{g^{2}}{\sqrt{2}}\frac{2}{3}%
\tan \theta _{w}N_{i1}^{\ast }U_{j1}^{\ast }(V_{CKM}^{\ast
})_{1a}(\delta
_{RL}^{d})_{2a}\right.\nonumber \\
&&\left. -\frac{g}{\sqrt{2}}\frac{2}{3}\tan \theta
_{w}N_{i1}^{\ast }U_{j2}^{\ast }(h_{d})_{33}(V_{CKM}^{\ast
})_{13}(\delta _{RR}^{d})_{23}\right),
\end{eqnarray}

\begin{eqnarray}
C_{3(\tau -\chi ^{0}-s)}^{q(1)} &=&(-\frac{1}{8})\left( \frac{2}{3}\frac{g^{2}}{\sqrt{2}}%
\tan \theta _{w}U_{j1}^{\ast }N_{i1}^{\ast }(V_{CKM}^{\ast
})_{1a}(\delta
_{RL}^{d})_{2a}\right.\nonumber \\
&&\left. -\frac{2}{3}\frac{g}{\sqrt{2}}\tan \theta
_{w}U_{j2}^{\ast} N_{i1}^{\ast }(V_{CKM}^{\ast })_{13}(
\delta_{RR}^{d})_{23}(h_{d})_{33}\right).
\end{eqnarray}

\be \tilde{C}_{4}=C_{4(\tau -\chi ^{-})}^{l(0)}C_{4(\tau -\chi
^{-}-u)}^{q(1)}\tilde{I}_{n}(x_{i},x_{j})+C_{4(\tau -\chi
^{0})}^{l(0)}C_{4((\tau -\chi
^{0}-u)}^{q(1)}\tilde{I}_{n}(x_{i},x_{j})+O(\delta ^{2}),
 \ee
where $\tilde{I}_{n}(x_{i},x_{j})$ is given below.

\begin{eqnarray}
C_{4(\tau -\chi ^{-})}^{l(0)}
&=&-(h_{e})_{33}\frac{g}{\sqrt{2}}(N_{i2}-\tan
\theta _{w}N_{i1})U_{j2}(U_{MNS}^{\ast })_{33} \nonumber\\
&=&C_{3(\tau -\chi ^{-})}^{l(0)},
\end{eqnarray}

\begin{eqnarray}
C_{4(\tau -\chi ^{0})}^{l(0)} &=&g(h_{e})_{33}N_{i3}U_{j1}^{\ast
}(U_{MNS}^{\ast
})_{33} \nonumber\\
&&-g\sqrt{2}\tan \theta _{w}N_{i1}U_{j2}^{\ast
}(h_{e})_{33}(U_{MNS}^{\ast })_{33},
 \end{eqnarray}

\begin{eqnarray}
C_{4(\tau -\chi ^{0}-u)}^{q(1)} &=&(-\frac{1}{8})\left( -\frac{4}{3}\frac{g^{2}}{\sqrt{2}}%
\tan \theta _{w}N_{i1}V_{j1}(V_{CKM}^{\ast })_{a2}(\delta
_{LR}^{u})_{a1}\right.  \nonumber\\
&&\left. +\frac{4}{3}\frac{g}{\sqrt{2}}\tan \theta
_{w}N_{i1}V_{j2}(h_{u})_{33}(V_{CKM}^{\ast })_{32}(\delta
_{RR}^{u})_{31}\right),
\end{eqnarray}%

\begin{eqnarray}
C_{4(\tau -\chi ^{-}-u)}^{q(1)} &=&(-\frac{1}{8})\left( -\frac{4}{3}\frac{g^{2}}{\sqrt{2}}%
\tan \theta _{w}V_{j1}N_{i1}(V_{CKM}^{\ast })_{a2}(\delta
_{LR}^{u})_{a1}\right.  \nonumber\\
&&\left. +\frac{4}{3}\frac{g}{\sqrt{2}}\tan \theta
_{w}V_{j2}N_{i1}(h_{u})_{33}(V_{CKM}^{\ast })_{32}(\delta
_{RR}^{u})_{31}\right).
\end{eqnarray}

 \be
\tilde{C}_{5(\tau -\chi ^{0}-u)}=-\frac{1}{4}\tilde{C}_{4(\tau
-\chi ^{0}-u)}, \ee

 \be
\tilde{C}_{5(\tau -\chi ^{-}-u)}=\frac{1}{4}\tilde{C}_{4(\tau
-\chi ^{-}-u)}. \ee

The loop integrals $I_{n}(x_{i},x_{j})$ and
$\tilde{I}_{n}(x_{i},x_{j})$ are defined as follows:

\begin{eqnarray}
I(x_{i},x_{j}) &=&\frac{1}{16\pi ^{2}\tilde{m}^{2}}\left( \frac{1}{%
x_{i}-x_{j}}\right) \left( \frac{x_{i}^{2}-x_{i}-x_{i}^{2}logx_{i}}{%
(1-x_{i})^{2}}-(x_{i}\leftrightarrow x_{j})\right) ,  \nonumber \\
\widetilde{I}(x_{i},x_{j}) &=&\frac{\sqrt{x_{i}x_{j}}}{16\pi ^{2}\tilde{m}%
^{2}}\left( \frac{1}{x_{i}-x_{j}}\right) \left( \frac{%
x_{i}^{2}-x_{i}-x_{i}logx_{i}}{(1-x_{i})^{2}}-(x_{i}\leftrightarrow
x_{j})\right) ,  \nonumber \\
I_{n}(x_{i},x_{j}) &=&\frac{1}{32\pi ^{2}\tilde{m}^{2}}\left( \frac{1}{%
x_{i}-x_{j}}\right) \left( \frac{2x_{i}^{2}-2x_{i}-2x_{i}logx_{i}}{%
(x_{i}-1)^{2}}-\frac{x_{i}^{3}-4x_{i}^{2}+3x_{i}+2x_{i}logx_{i}}{%
(x_{i}-1)^{3}}\right.\nonumber\\
&-&\left.(x_{i}\leftrightarrow x_{j})\right) ,~~~  \nonumber \\
\tilde{I_{n}}(x_{i},x_{j}) &=&\frac{-\sqrt{x_{i}x_{j}}}{32\pi ^{2}\tilde{m}%
^{2}}\left( \frac{1}{x_{i}-x_{j}}\right) \left( \frac{%
x_{i}^{3}-4x_{i}^{2}+3x_{i}+2x_{i}logx_{i}}{(x_{i}-1)^{3}}%
-(x_{i}\leftrightarrow x_{j})\right) .
\end{eqnarray}

%%%%%%%%%%%%%%%%%%%%%%%%%%%%%%%%%%%%%%%%%%

\end{document}